%% file: 0_main.tex
\documentclass{article}  

\usepackage{0_arxiv}

\usepackage[utf8]{inputenc} 
\usepackage[T1]{fontenc}    
\usepackage{hyperref}       
\usepackage{url}            
\usepackage{booktabs}       
\usepackage{wrapfig}        
\usepackage{amsmath}
\usepackage{amsfonts}       
\usepackage{mathrsfs}
\usepackage{mathtools}
\usepackage{dsfont}
\usepackage{nicefrac}       
\usepackage{microtype}      
\usepackage{todonotes}
\usepackage{accents}        
\usepackage{listings}
\usepackage{svg}
\definecolor{codegreen}{rgb}{0,0.6,0.05}
\definecolor{codegray}{rgb}{0.5,0.5,0.6}
\definecolor{codepurple}{rgb}{0.58,0,0.82}
\definecolor{backcolour}{rgb}{1, 1, 1}
\lstdefinestyle{python_sytle}{
    inputencoding=latin1,
    backgroundcolor=\color{backcolour},   
    commentstyle=\color{codegreen},
    keywordstyle=\color{magenta},
    numberstyle=\tiny\color{codegray},
    stringstyle=\color{codepurple},
    basicstyle=\footnotesize\selectfont\ttfamily,
    breakatwhitespace=false,         
    breaklines=true,                 
    captionpos=b,                    
    keepspaces=true,                 
    numbers=left,                    
    numbersep=3pt,                  
    showspaces=false,                
    showstringspaces=false,
    showtabs=false,                  
    tabsize=2
}

\newlength{\dhatheight}     

\newcommand{\x}{\mathbf{x}}

\title{TMM-Fast: A Transfer Matrix Computation Package for Multilayer Thin-Film Optimization}

\author{
        Alexander Luce \\
        Max Planck Institute for the Science of Light\\
        Friedrich-Alexander-Universität\\
        Erlangen-Nürnberg\\
        OSRAM Opto Semiconductors GmbH\\
        Regensburg \\
        \texttt{alexander.luce@ams-osram.com} \\
    \And
        Ali Mahdavi \\
        OSRAM Opto Semiconductors GmbH\\
        Regensburg \\
    \AND
        Florian Marquardt \\
        Max Planck Institute for the Science of Light\\
        Friedrich-Alexander-Universität\\
        Erlangen-Nürnberg\\ 
    \And
        Heribert Wankerl \\
        Universität Regensburg\\
        OSRAM Opto Semiconductors GmbH\\
        Regensburg \\
}

\begin{document}
\maketitle

\begin{abstract}
\input{1_abstract}
\end{abstract}


\section{Introduction}
\input{1_introduction}

\section{Physical Background of the Transfer Matrix Method (TMM)}
\label{sec:transfer_matrix_method}
\input{2_transfer_matrix_method}


\section{Implementation in Numpy}
\label{sec:implementation}
\input{3_implementaion_in_numpy}

\section{TMM-Torch: Multilayer thin-film gradients via Autograd}
\label{sec:tmm_torch}
\input{4_tmm_torch}

\section{Environment for reinforcement learning}
\label{sec:environment}
\input{5_Environment_for_reinforcement_learning}

\section{Conclusion}
\label{sec:conclusion}
\input{6_conclusion}

\bibliographystyle{unsrt}  
\bibliography{0_references}  

\appendix
\section{Appendix}
\label{sec:appendix}
\input{7_appendix}







\end{document}

%% file: 1_abstract.tex

Achieving the desired optical response from a multilayer thin-film structure over a broad range of wavelengths and angles of incidence can be challenging. An advanced thin-film structure can consist of multiple materials with different thicknesses and numerous layers. Design and optimization of complex thin-film structures with multiple variables is a computationally heavy problem that is still under active research. To enable fast and easy experimentation with new optimization techniques, we propose the Python package TMM-Fast which enables parallelized computation of reflection and transmission of light at different angles of incidence and wavelengths through the multilayer thin-film. By decreasing computational time, generating datasets for machine learning becomes feasible and evolutionary optimization can be used effectively. Additionally, the sub-package TMM-Torch allows to directly compute analytical gradients for local optimization by using PyTorch Autograd functionality. Finally, an OpenAi Gym environment is presented which allows the user to train reinforcement learning agents on the problem of finding multilayer thin-film configurations.


%% file: 1_introduction.tex
\begin{figure}[!ht]
\centering
    \includegraphics[width=1.\textwidth, trim = 1.5cm 18.5cm .7cm .5cm, clip]{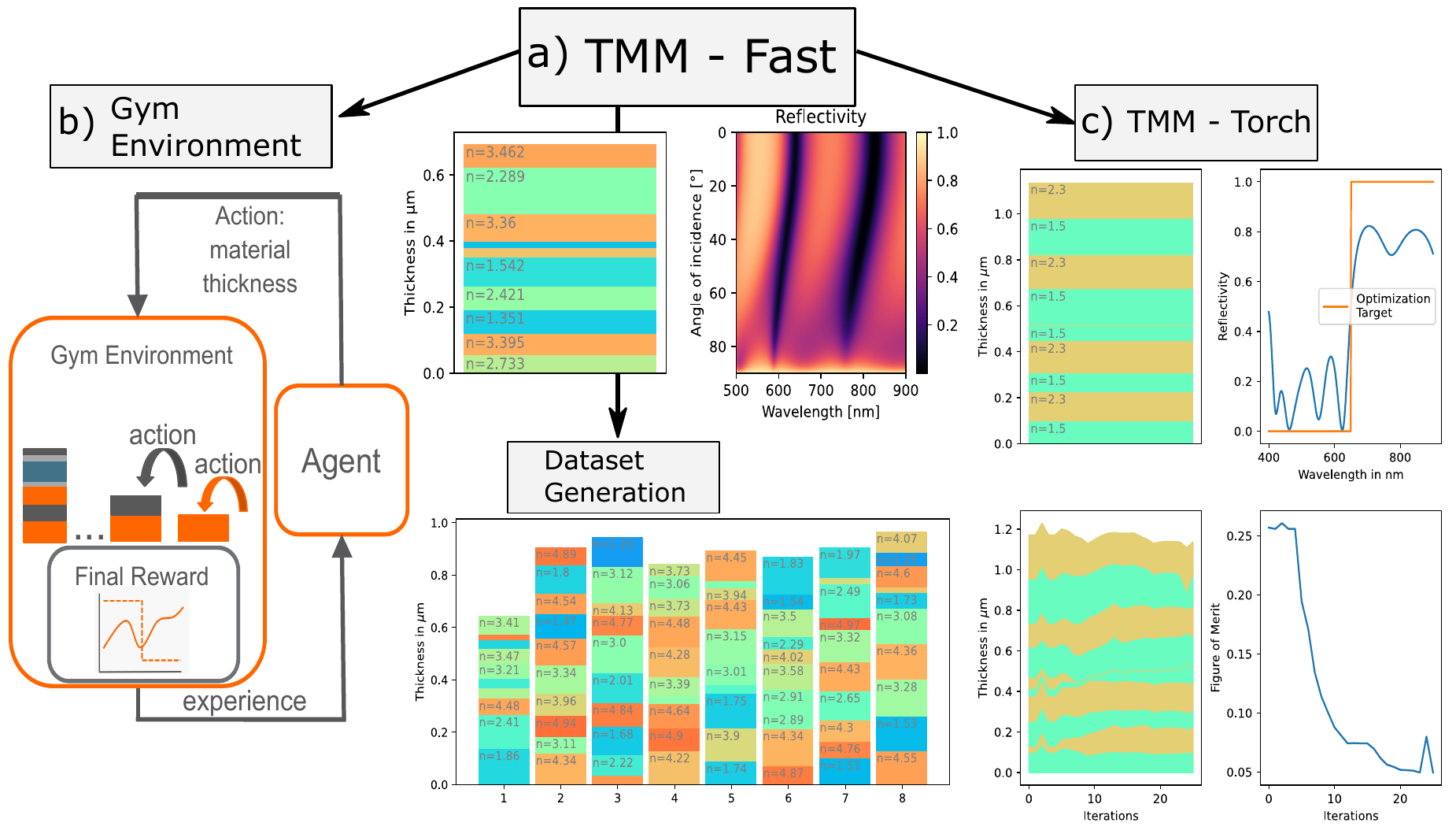} 
\caption{Schematic overview of the contents of the TMM-Fast package. The package consists of three subdivisions. Part a) of \autoref{fig:tmm_package_overview} contains the core functionality of the TMM-package, computing the optical response of a multilayer thin-films quickly via the transfer matrix method over a broad range of wavelengths and incident angles. Another part of the core functionality of solving multilayer thin-films is the possibility to generate huge datasets with >1e5 data samples for machine learning models. \\
Part b) encompasses an OpenAI Gym Environment which allows easy comparisons of different reinforcement learning agents. Here, the environment state is given by the total layer number, layer thicknesses, and material choice. The agent takes an action that specifies the material and thickness of the layer to stack next. The environment implements the multilayer thin-film generation as consecutive conduction of actions and assigns a reward to a proposed multi-layer thin film based on how close the actual (solid orange line) fulfils a desired (dashed orange line) characteristic - in this case, a one-dimensional optical characteristic, e.g. reflectivity over wavelength for perpendicular incidence. The experience accumulated by the taken actions of the agent is used to adapt subsequent actions in order to increase the reward and thus generate more and more sophisticated multilayer thin-films. An investigation of a reinforcement learning agent which was trained with the proposed Gym Environment is given by Wankerl et. al. \cite{wankerl2020parameterized}. \\
Part c) encompasses an implementation of the transfer matrix method via PyTorch functions which allow for backpropagation through the entire computation. This enables easy differentiation of the parameters of interest and allows gradient based and quasi-newton optimization methods to be used for optimization of multilayer thin-films. In the example shown, a random thin-film is optimized with respect to the given optimization target, weighted by the mean squared error. Gradients via autograd are used with the BFGS algorithm \cite{Lecun98gradient-basedlearning}. The iterative evolution of the layer thicknesses and the correspondingly decreasing figure of merit are shown below. For the given initial values the optimization converged in about 25 iteration steps.}
\label{fig:tmm_package_overview}
\end{figure}

Although the existence of a globally optimal multilayer thin-film was proven \cite{Tikhonravov1993,Tikhonravov1993b, Ebrahimi2018}, the optimization of multilayer thin-films concerning reflectivity and transmittivity over wavelength and incidence angle remains a challenging task for the scientific community \cite{Anzengruber2012, Becker2014, Liddell1981}. As such it gained attention in many publications \cite{Chang1990, Paszkowicz2013, Yang2013, Guo2014, Martin1995} including various methods based on reinforcement learning \cite{Jiang2020, wankerl2020parameterized} and machine learning \cite{Hedge2019, Roberts2018}. Naturally, the need for a standardized, encompassing environment arises to make future research more trustful, comparable, and consistent. Here, we propose a comprehensive Python package that provides functionality to researchers to design and optimize multilayer thin-films. At its core, the proposed TMM-Fast package (Transfer Matrix Method - Fast) is a parallelized and re-implemented revision of the existing TMM code that was initially published by S. Byrnes \cite{byrnes2019multilayer}. It implements Abèles transfer matrix method \cite{Abels1950LaTG} in Python to calculate transmission and reflection of an incident plane wave through a slab of layered thin-film materials with different thicknesses. Given a broad, discretized spectrum of light irradiating a thin-film under particular incident angles, the transmission and reflection must be calculated for each contributing wavelength at each angle of incident. The coefficients of transmission and reflection for a specific wavelength and angle of incidence depend on the thicknesses and dispersive and dissipative refractive index of the layers. Those coefficients can deviate significantly at wavelengths which differ only slightly for the same multilayer thin-film. An intuitive approach to increase the computational speed of the response of a multilayer thin-film is to parallelize the pair-wise independent computations regarding wavelengths and angles. The parallelization is implemented via matrix operations based on the open-source numeric algebra package Numpy \cite{numpy} and the thread-management package Dask. The parallelization showed to reduce the computational time by $\sim 100\times$ with an additional factor on the order of available CPU cores by using Dask. Additionally, by using PyTorch Autograd \cite{Paszke2017AutomaticDI}, analytical gradients of the multilayer thin-film layer thicknesses and refractive indices can be computed for local optimization. Since the implementation is done via PyTorch methods, it can be integrated into advanced Neural Networks. Moreover, the parallelized TMM-Fast code is used to implement an OpenAI Gym environment \footnote{\url{https://github.com/MLResearchAtOSRAM/gym-multilayerthinfilm}} that can be used by physicists and reinforcement learning researchers. In the environment, the optimization of multilayer thin-films is introduced as a sequence generation process and thus considered as a parameterized Markov decision process \cite{Masson2016}, shown in section \ref{sec:environment}. An overview of the scope of the package is shown in \autoref{fig:tmm_package_overview}.
All code is available on GitHub\footnote{\url{https://github.com/MLResearchAtOSRAM/tmm_fast}} under MIT license.

\vspace{0.5cm}

%% file: 2_transfer_matrix_method.tex
Light of a particular wavelength passing from one into another material experiences a sudden change of the refractive index $n_1$ to $n_2$ that results in reflection and transmission of the incoming wave. A light wave with a wavevector $\textbf{k}$ which is not parallel to the surface normal experiences refraction. 
The ratio of angle of incidence $\theta_1$ and angle of refraction $\theta_2$ is given by Snell's law
    $\frac{\sin{\theta_1}}{\sin{\theta_2}} = \frac{n_2}{n_1}$.
Given the Fresnel equations, reflection and transmission coefficients $r$ and $t$ can be computed, where a distinction is made between s and p polarization by using a subscript
\begin{align}
    t_s = \frac{2n_1 \cos{\theta_1}}{n_1\cos{\theta_1} + \frac{\mu_{r1}}{\mu_{r2}}n_2\cos{\theta_2}} \hspace{0.5cm}
    r_s = \frac{n_1\cos{\theta_1} - \frac{\mu_{r1}}{\mu_{r2}}n_2\cos{\theta_2}}{n_1\cos{\theta_1} + \frac{\mu_{r1}}{\mu_{r2}}n_2\cos{\theta_2}}\\
    t_p = \frac{2n_1 \cos{\theta_1}}{\frac{\mu_{r1}}{\mu_{r2}}n_2\cos{\theta_1} + n_1\cos{\theta_2}} \hspace{0.5cm}
    r_p = \frac{\frac{\mu_{r1}}{\mu_{r2}}n_2\cos{\theta_1} - n_1\cos{\theta_2}}{\frac{\mu_{r1}}{\mu_{r2}}n_2\cos{\theta_1} + n_1\cos{\theta_2}}
\end{align} 
and $\mu_{r}$ being the respective magnetic permeability. Note that for $\theta_1$ = 0$^\circ$, i.e. vertical incidence, $r_s = r_p$ and $t_s = t_p$. Based on the reflection and transmission coefficients, the reflectivity and transmittivity R and T can be computed as shown in \autoref{eq:transmittivity}. 
\begin{align}
    R_{\lambda, \theta_1}= r_i^2; \hspace{0.5cm} 
    T_{\lambda, \theta_1} = \frac{n_2\cos{\theta_2}}{n_1\cos{\theta_1}}t_i^2 \label{eq:transmittivity}
\end{align}
where the subscript $i$ indicates the polarization. Note that, for absorptionless materials $T+R=1$ holds.

Now, consider a multilayer thin-film with $L\in\mathds{N}$ layers and the individual layers are denoted by $l \leq L$. The light enters from an injection layer of semi-infinite thickness $l=0$ with a relative amplitude of 1 and exits the multilayer thin-film in the outcoupling layer of semi-infinite thickness with $l=L+1$. From the outcoupling layer, no light enters the thin-film. The transmitted part of the light in layer $l$, that travels in "forward" direction, ie. towards the layer with $l=l_{\text{current}\, +\,1}$ is denoted by $v_l$ and the reflected part that travels in "backward" direction is given by $w_l$. By using the reflection and transmission coefficients r and t, the response of any layer can be written as
\begin{align}
    \begin{pmatrix}
        v_l\\
        w_l
    \end{pmatrix} = 
    \begin{pmatrix}
        e^{-i\delta_l} & 0 \\
        0 & e^{i\delta_l}
    \end{pmatrix}
    \begin{pmatrix}
        1 & r_{l, l+1} \\
        r_{l, l+1} & 1 
    \end{pmatrix}
    \frac{1}{t_{l, l+1}}
    \begin{pmatrix}
        v_{l+1}\\
        w_{l+1}
    \end{pmatrix}= M_l 
    \begin{pmatrix}
        v_{l+1}\\
        w_{l+1}
    \end{pmatrix}.
\end{align}
where $\delta=d_l\,k_z$ is the accumulated phase of the light wave when travelling through a layer with a specific thickness $d_l$ and with wave vector $k_l$.

Eventually, the total characteristic matrix of the multilayer thin-film is given by
\begin{align}\label{eq:m_tilde}
    \Tilde{M} = \prod_{i=0}^{L-1} M_i.
\end{align}
Finally, to compute the reflection and transmission of the entire multilayer thin-film, one needs to evaluate
\begin{align}
    \begin{pmatrix}
        1\\
        r
    \end{pmatrix} = \Tilde{M}
    \begin{pmatrix}
        t\\
        0
    \end{pmatrix}.
\end{align}
The transmission and reflection coefficients are separated
\begin{align}
    r_{\lambda, \vartheta} = \frac{\Tilde{M}_{10}}{\Tilde{M}_{00}} \hspace{1cm} t_{\lambda, \vartheta} = \frac{1}{\Tilde{M}_{00}}.
\end{align}
and allow to compute the reflectivity and transmittivity via \autoref{eq:transmittivity}.
Note that one can easily calculate the partial transmission and reflection coefficients by only multiplying \autoref{eq:m_tilde} up to $L-l$. The initial angle of incidence in the first layer is given by $\vartheta$.

%% file: 3_implementaion_in_numpy.tex
The key contribution of TMM-Fast is the parallelized handling of the characteristic matrix that reduces computational time. The matrix $M_l$ consists of three separate matrices, matrix $A$ which encompasses the accumulated phase, and the two matrices holding the coefficients of reflection and transmission, respectively. They are of shape $\left[N_\lambda, N_\vartheta, L, 2, 2\right]$, where $N_\lambda$ and $N_\theta$ are the number of wavelengths and incident angles, respectively. To get the characteristic matrix $M_l$, Numpy's \textit{einsum}\footnote{Documentation: \url{https://numpy.org/doc/stable/reference/generated/numpy.einsum.html}} method allows to specify multiplication and contractions of different dimensions easily:
\begin{lstlisting}[language=Python, style = python_sytle]
M_l = np.zeros((num_lambda, num_angles, num_layers, 2, 2), 
                   dtype=complex)
F = r_list[:, 1:]
M_l[:,:,1:-1,0,0] = np.einsum('hji,ji->jhi', 1/A, 1/t_list[:, 1:])   
M_l[:,:,1:-1,0,1] = np.einsum('hji,ji->jhi', 1/A, F/t_list[:, 1:]) 
M_l[:,:,1:-1,1,0] = np.einsum('hji,ji->jhi', A, F/t_list[:, 1:])  
M_l[:,:,1:-1,1,1] = np.einsum('hji,ji->jhi', A, 1/t_list[:, 1:]) 

Mtilde = np.empty((num_angles, num_lambda, 2, 2), dtype=complex)
Mtilde[:, :] = make_2x2_array(1, 0, 0, 1, dtype=complex)
for i in range(1, num_layers-1):
        Mtilde = np.einsum('ijkl,ijlm->ijkm', Mtilde, M_l[:,:,i])
\end{lstlisting}
Finally $\Tilde{M}$ is computed by multiplying out the thickness dimensions. 

The entire function is called coh\_tmm\_fast or by coh\_tmm\_fast\_disp. Both methods differ since the former assumes dispersionless materials whereas the latter accepts dispersive materials. An Example is give in Appendix \ref{sec:appendix-core}.

\section{Core functionality speedup and dataset generation}
To verify the speedup that the TMM-Fast package provides compared to the native implementation of Byrnes, ten multilayer thin-films are generated with 21 layers are generated and evaluated in a spectral range from 400 nm to 700 nm at 100 equally spaced points. The angles of incidence range from 0° to 90° based on 20 equally spaced supporting points. The original TMM method requires a computation time of $17.5 \pm 0.131$ s for the evaluation while our proposed method computes the coefficients of the multilayer thin-film in $0.168 \pm 0.002$ s which corresponds to an acceleration of $\sim$ 100$\times$. 

Finally, the Python package Dask\footnote{\url{https://dask.org/}} manages parallel threads of the CPU to distribute the computation of different independent computations on all available CPU cores. The application is straightforward: by calling the \textit{coh\_tmm\_fast} function implicitly inside the \textit{delayed()} function of Dask, a list of all necessary computations is created. Then, the entire list is executed implicitly to create a computational graph for Dask, which is required to orchestrate the parallel threads efficiently. Lastly, the \textit{compute()} method triggers the actual computation and the result is returned. By running coh\_tmm\_fast on all available threads, an additional speedup of the dataset creation on the order of the number of available CPU cores is possible. However, the benefit might decrease for very large computational clusters with many cores since the management of the parallel threads creates computational overhang. By calling \textit{multithread\_coh\_tmm} from the TMM-Fast package, the computation is easily started. A Dataset with one million samples of 9-layer multilayer thin-films at 100 wavelengths and 10 angles of incidence can be created in approximately 40 min on an eight-core machine. Computing more layers, wavelengths points, and angles can increase the computational time significantly. An example is shown in the Appendix \autoref{sec:dataset_generation}.

%% file: 4_tmm_torch.tex
For optimization, computing gradients enable gradient-based optimization algorithms such as gradient descent \cite{Lecun98gradient-basedlearning}. These gradient based optimization methods generally converge to local minima with fewer iterations than other non-gradient based algorithms such as the Nelder-Mead downhill simplex algorithm \cite{Nelder1965ASM}. The Python package PyTorch\footnote{\url{https://PyTorch.org/}} implements matrix multiplication methods which allow parameters to be automatically differentiated (\textit{Autograd}) via the chain rule \cite{Paszke2017AutomaticDI}. Autograd enables the user to compute the gradients of the input parameters without the necessity of deriving the gradient analytically and can be dynamically adapted to the problem. By using the TMM-Torch subpackage of TMM-Fast, the gradients for a multilayer thin-film can be readily computed. An example is shown in the Appendix \ref{sec:appendix-torch}. The application of the transfer matrix takes slightly longer by using the PyTorch routines. Therefore, using the regular TMM-Fast algorithms is still advised to reduce computational time if gradients are not necessary.

%% file: 5_Environment_for_reinforcement_learning.tex

Reinforcement learning \cite{Sutton1998} is an area of machine learning concerned with how intelligent agents ought to take actions in an environment in order to maximize a notion of reward. The proposed code implements such an environment, where agents can stack, characterize and optimize multilayer thin-films. Therefore, the generation of multilayer thin-films is considered as a parameterized Markov decision processes \cite{Masson2016} and thereby implemented as a sequence generation process: Beginning from $l=1$, an agent subsequently executes parameterized actions $a_l = \left( d_l, m_l \right)$ that specify the thickness $d_l$ and material index $m_l$ of the $l$-th layer. These actions determine which material of which thickness to stack next, thereby consecutively forming a multilayer thin-film as illustrated in \autoref{fig:tmm_package_overview}. The stacked $l$ layers and - optional - the optical characteristics of these intermediate multilayer thin films are provided to the agent as the environmental state $s$. Given this state, the agent takes the next action $a_{l+1}$ until the pre-defined maximum number of layers is reached or the agent decides to terminate stacking. The optical characteristic of the final proposed multilayer thin-film of $L$ layers, e.g. regarding reflectivity $R_{ \lambda, \vartheta} \left( \mathbf{I},\mathbf{d}\right)$ over wavelength $\lambda$ and angle $\vartheta$ of incidence, is computed via the proposed transfer-matrix method (TMM-Fast). Here, $\mathbf{I}\in \mathbb{C}^{L \times N_{\lambda}}$ refers to the matrix of (dispersive and dissipative) refractive indices of the material of each layer. Each material is identified by the material identifier index $m_l$. $\mathbf{d} \in \mathbb{R}^L$ denotes the vector of layer thicknesses. When the agent starts to stack a thin-film, $\mathbf{d}$ and $\mathbf{I}$ are initialized with zeros and get filled according to the taken actions with the layer thicknesses and (dispersive) refractive indices, respectively. The observed reflectivity is compared to a user-defined, desired reflectivity $R^{\text{target}}_{ \lambda, \vartheta}$, in order to derive a notion of numeric reward, e.g. an inverted reconstruction error 
$$
-\sum_{\lambda, \vartheta} \vert R^{\text{target}}_{ \lambda, \vartheta}
-
R_{ \lambda, \vartheta}\left( \mathbf{I}, \mathbf{d} \right)
\vert.
$$ Based on this reward, the agent learns --- for example, based on Q-learning \cite{Jiang2020, Watkins1989} --- to distinguish between good and bad actions and thus derive an optimal thin-film design.

Whereas the contained physical methods are well-studied and known for decades, the contribution of the code available at \url{https://github.com/MLResearchAtOSRAM/gym-multilayerthinfilm} lies in the implementation of an OpenAI gym-related environment. Here, the intention is to enable AI researchers without optical expertise to solve the corresponding parameterized Markov decision processes based on common code in the future.

%% file: 6_conclusion.tex
In this technical note, the comprehensive Python package TMM-Fast for multilayer thin-film computation, optimization and reinforcement learning is presented. At its core, the package is comprised of revised and speedup transfer matrix method code from the original TMM package\cite{byrnes2019multilayer}. TMM-Fast enables the user to compute multilayer thin-films with Numpy and PyTorch methods and gives full control over the data to the user. Since the TMM-Fast package evaluates multilayer thin-films especially fast, it can be used to generate datasets for machine learning. The reduced computational time also enables evolutionary optimization to be executed in a reasonable amount of time. The TMM-Torch implementation allows the user to compute analytical gradients via automatic differentiation. Quasi-Newton or gradient based optimization algorithms such as gradient descent can be used and converge faster to local minima by using an analytical gradient. Finally, an OpenAi Gym environment is proposed which allows researchers to easily test and experiment with new reinforcement agents on solving the multilayer thin-film problem. All code proposed in this paper is open-source and available at \url{https://github.com/MLResearchAtOSRAM/} under the MIT-licence. 

%% file: 7_appendix.tex
\subsection{TMM-Fast Example}\label{sec:appendix-core}
To demonstrate the functionality, a minimal example is given for a multilayer thin-film with random thicknesses and refractive indices. 

\begin{figure}[ht]
\centering
    \includegraphics[angle=0,trim = .2cm .2cm .2cm 0.cm, clip, width=1.\textwidth]{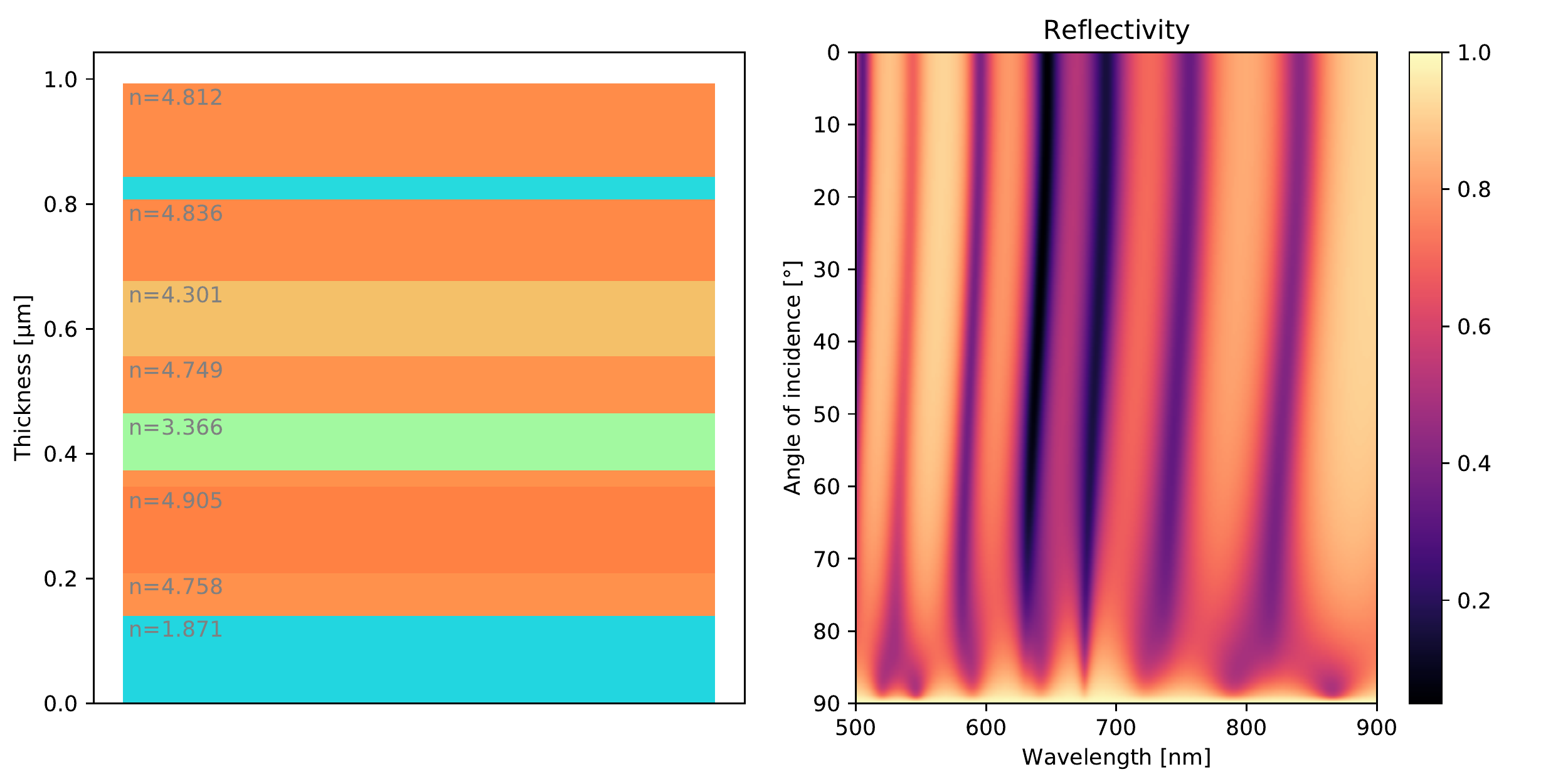}\quad 
\caption{A multilayer thin-film with random layer thicknesses and refractive index under unpolarized illumination. In this example, the materials are dispersion- and dissipationless. The injection region, which is below the thin-film in this depiction, possesses a refractive index of $n=2$. The outcoupling region above the thin-film possesses a refractive index of $n=1$. The reflectivity is computed over a wavelength range of 500 nm to 900 nm and from 0$^\circ$ to 90$^\circ$ on a 300$\times$300 grid. \label{fig:example_reflectivity}}
\end{figure}

\begin{lstlisting}[language=Python, style = python_sytle]
import tmm_fast as tmmf
L = 12 # number of layers
d = np.random.uniform(20, 150, L)*1e-9 # thicknesses of the layers
d[0] = d[-1] = np.inf # set first and last layer as injection layer
n = np.random.uniform(1.2, 5, L) # random constant refractive index
n[-1] = 1 # outcoupling into air

wl = np.linspace(500, 900, 301)*1e-9
theta = np.deg2rad(np.linspace(0, 90, 301))
# here s and p polarization is computed and averaged to simulate incoherent light
result = (tmmf.coh_tmm_fast('s', n, d, theta, wl)['R'] 
                + tmmf.coh_tmm_fast('p', n, d, theta, wl)['R'])/2
\end{lstlisting}
The result of the computation can then be further evaluated. By using the \textit{plot\_stacks} function from the TMM-Fast package, the multilayer thin-film can be visualised. An example is shown in \autoref{fig:example_reflectivity}.

\subsection{Dataset Generation}\label{sec:dataset_generation}
A dataset containing tens of thousands of datasamples can be created by using the \textit{multithread\_coh\_tmm} function. The external package \textit{tqdm} is used to display a progress bar in order to keep track of the generation progress. In this example, a dataset with 1e6 thin-films at 100 wavelengths should be created for vertical incidence.
\begin{lstlisting}[language=Python, style = python_sytle]
import numpy as np
import tqdm

n_samples = 1e6
n_lambda = 100
n_layers = 12
wavelength = np.linspace(1000, 1700, n_lambda)*1e-9

theta = np.array([0]) # vertical incidence
stack_layers = np.random.uniform(5, 180, (n_samples, n_layers))*1e-9
stack_layers[:,0] = stack_layers[:,-1] = np.inf # injection and outcoupling layer
optical_index = np.array([2.5] +[2.0, 1.4]*5 +[1.])
optical_index = np.tile(optical_index, (n_samples, 1))
n = 10000 # the dataset is computed in steps of 1e5
dataset = np.empty((n_samples, n_lambda))
for i in tqdm.tqdm(np.array(range(n_samples))[::n]):
    tmm_res = np.empty((n, n_lambda))
    tmm_res = multithread_coh_tmm('s', optical_index[i:i+n], stack_layers[i:i+n], theta, wavelength, TorR='R').squeeze()
    dataset[i:i+n] = tmm_res
\end{lstlisting}

Now the dataset can be saved for example by using .hdf or Numpy's .npz format.

\subsection{TMM-Torch Example}\label{sec:appendix-torch}
Here, an example how to compute gradients via automatic differentiation with the TMM-Torch package is shown.

\begin{lstlisting}[language=Python, style = python_sytle]
import torch
import tmm_fast_torch as tmmt
n_layers = 12 # number of layers
stack_layers = np.random.uniform(20, 150, n_layers)*1e-9 # thicknesses of the layers
stack_layers[0] = stack_layers[-1] = np.inf # set first and last layer as injection layer
optical_index = np.random.uniform(1.2, 5, n_layers) # random constant refractive index
optical_index[-1] = 1 # outcoupling into air

stack_layers = torch.tensor(stack_layers, requires_grad=True)

wl = np.linspace(500, 900, 301)*1e-9
theta = np.deg2rad(np.linspace(0, 90, 301))

result = tmmt.coh_tmm_fast('s', optical_index, stack_layers, theta, wavelength)['R']
mse = torch.nn.MSELoss()
error = mse(result, torch.zeros_like(result)
error.backward()

gradients = stack_layers.grad
\end{lstlisting}
Since the error is computed with respect to zero reflectivity for all wavelengths at all incidences, the gradient points towards the steepest descent for a broadband anti-reflection coating. By using the minimize function from the Python package Scipy and a gradient based optimization algorithm, for example "L-BFGS-B"\footnote{\url{https://docs.scipy.org/doc/scipy/reference/optimize.minimize-lbfgsb.html}}, a local optimum is easily found.